\documentstyle[12pt]{article}
\newcommand{\bb}{\begin{eqnarray}}
\newcommand{\ee}{\end{eqnarray}}

\newcommand{\m}{\mu}
\newcommand{\n}{\nu}

\newcommand{\pl}{\partial}

\begin{document}
\begin{titlepage}
\begin{center}
\vspace*{.2cm}
\hspace*{10cm} TUM-HEP-238/96\\
\hspace{10cm} hep-th/9602059\\
\hspace{10cm} February 1996\\
\vspace*{1.5cm}

{\LARGE Type IIA/IIB string duality for targets with abelian isometries}\\

\vspace{1cm}

{\large Alexandros A. Kehagias}
\footnote{Supported by an Alexander von Humboldt-Stiftung}\\
\vspace{-.2cm}
{\large Physik Department \\
\vspace{-.2cm}
Technische Universit\"at M\"unchen\\
\vspace{-.2cm}
D-85748 Garching, Germany\\
E-Mail: kehagias@physik.tu-muenchen.de}\\
\end{center}
\vspace{1.3cm}

\begin{center}
{\large Abstract}
\end{center}
\vspace{.3cm}

We examine T-duality transformations for supersymmetric strings with target
space geometry  with compact
abelian isometries. We consider the partition function of
these models and we show that although T-duality is not a symmetry, due to an
anomaly, it
relates type IIA to type IIB strings. In this way we extend the corresponding
result for toroidal compactification to the general case
of non-trivial backgrounds with abelian isometries and
for world sheets of any genera.
\end{titlepage}
\newpage

In a previous work \cite{1}
 we have examined  the transformation properties of the
$\sigma$-model
 partition function under T-duality \cite{2}--\cite{2''}.
In particular, we have considered
a N=1 supersymmetric 2D $\sigma$-model with
target space a manifold M with a compact
abelian isometry. For this model we have
written down the corresponding partition function $Z[g,b,s]$ which in general
depends on the metric $g_{\m\n}$ and the antisymmetric field $b_{\m\n}$ of M,
 as well as on the spin structure ${\it s}$ of the world sheet. By performing
 the standard duality transformation for targets  with abelian isometries
 \cite{bus}, we found
that, besides the well-known transformations of the target metric, the
antisymmetric field and the
dilaton shift \cite{dil1},\cite{dil2},
there also exists a transformation of the fermion
fields \cite{1},\cite{bus},\cite{2''}.
In particular, left and right-handed fermions are transformed
differently under duality giving  rise to a corresponding non-trivial
 transformation of
the fermionic measure. We have calculated the Jacobian of this transformation
using Fujikawa's method \cite{F} and we have  found
an ``anomaly" which depends  on the parity
of the spin structure of the world sheet \cite{1}.
As a result, T-duality, due to this anomaly,  is not
strictly speaking a symmetry of the N=1 supersymmetric $\sigma$-model.
Here we will show that for the string case, this  anomaly  is what
one needs to establish the type IIA/IIB string duality in the general case.

That the type IIA theory
is perturbatively T-dual to type IIB  has long been known
for the 10-dimensional flat Minkowski space compactified on an odd dimensional
torus \cite{dine}.
 The crucial observation here was that besides the $R\rightarrow 1/R$
transformation, one has to transform the fermionic
left movers as well in order to preserve
supersymmetry. For compactification on a circle in the ninth direction for
example, the
fermionic transformation is $\psi^9_L\rightarrow -\psi^9_L$. This
transformation flips the sign of $\Gamma^{11}_L$, alters the GSO projection and
change the type IIA theory to type IIB and vice versa. We will
extend here this result for non-trivial backgrounds with compact abelian
isometries and for world sheets of any genera.

Let us consider a $N=1$ supersymmetric $\sigma$-model defined on a 2-dim
space-time $\Sigma$ with metric of $(-,+)$ signature.
The target space M is parametrized by the scalars $(X^I,I,J=1,\cdots,D)$, its
metric is $g_{IJ}$  and the  antisymmetric field $b_{IJ}$.
The  action for this model, in the
conformal gauge and in the conventions of \cite{vp} is
\bb
S=\frac{1}{2} \int \hspace{-.3cm}&d^2\sigma\hspace{-.3cm}
&\left( (g_{IJ}+B_{IJ})\pl_+ X^I\pl_-X^J
-ig_{IJ}\psi^I_+D_+ \psi^J_+\right. \nonumber \\
&&\left.-ig_{IJ}\psi^I_-D_- \psi^J_-
+\frac{1}{2}\psi^I_+\psi^J_+\psi^K_-\psi^L_-R_{IJKL}(\Gamma_-) \right)
\, , \label{saa}
\ee
where
\bb
D_\pm\psi^I_\pm=\partial_\mp\psi^I+({\Gamma^I}_{KL}\pm \frac{1}{2}
 {H^I}_{KL})\partial_\mp
X^K\psi^L_\pm \, ,\label{psi}
\ee
with $H_{IJK}=\partial_I b_{JK}+cycl.\, perm.$ the torsion and the Riemann
tensor is evaluated with the torsionful connection
${\Gamma^I}_{JK}+\frac{1}{2}{H^I}_{JK}$.
 The action (\ref{saa}) is invariant under the supersymmetry transformations
\bb
\delta_\pm X^I&=&\mp i\epsilon_\pm \psi^I_\mp\,  , \nonumber\\
\delta_\pm\psi^I_\mp&=&\mp\partial_\mp X^I\epsilon_\pm \, , \nonumber \\
\delta_\pm\psi^I_\pm&=&\pm i\psi^J_\pm\epsilon_\pm({\Gamma^I}_{JK}
\mp \frac{1}{2} {H^I}_{JK})\psi^K_\mp \, , \label{sup}
\ee
as well as under reparametrizations (diffeomorphisms) of $M$
\bb
{X^I}^\prime={X^I}^\prime(X^J)&,& {\psi^I}^\prime=\frac{\pl {X^I}^\prime}
{\pl X^J}\psi^J \, , \label{diff}
\ee
which, moreover, commute with the supersymmetry transformations
of eq.(\ref{sup}).  The  partition function of the theory is obtained by
integrating  over all $(X^I,\psi^I)$ fields and it is given by
\bb
Z[g,b,s]=\int d\m_bd\m_f e^{-iS}=
\int [\sqrt{g}{\cal{DX}}^I][\frac{1}{\sqrt{g}}
{\cal D}\psi^I]e^{-iS}\, ,\label{pf}
\ee
where $g=det(g_{IJ})$ is the determinant of the target-space metric.
It depends, in general, on the background metric
and antisymmetric field and moreover on
the spin structure $s$ of the world sheet.
Analytic continuation to Euclidean time is necessary for a proper definition of
$Z[g,b,s]$. However, such a continuation can be
performed  after  the duality transformation.
We have not considered ghosts contributions
because the ghost sector is not
affected by the presence of the background metric \cite{ghost}
and thus, it is irrelevant for our purposes.

 We will assume now, that $M$ has a compact  abelian isometry,
orthogonal, for simplicity, to the surfaces of transitivity, so that
the target space metric may  be written as
\bb
g_{IJ}(X^K)=(g_{ij}(X^k),g_{00}(X^k))\, , \label{met}
\ee
where $X^K=(X^k,X^0=X^D)\, , (i,j,k=1,\cdots,D-1)$
are  coordinates adapted to the congruence  $\pl/\pl X^0$.
The latter generates  an  isometry which
corresponds to a $X^0\rightarrow
X^0+const.$ symmetry. The duality transformation is performed by gauging this
symmetry and adding a Lagrange multiplier which constraints the field strength
to vanish. The integration of the multiplier gives back the original model
while by integrating out the gauge field, the dual model is obtained.
 Applying this procedure to the present case, we get  the dual action
\bb
\tilde{S}=\frac{1}{2\pi} \int \hspace{-.3cm}&d^2\sigma\hspace{-.3cm}
&\left( (\tilde{g}_{IJ}+\tilde{b}_{JK})
\pl_+ X^I\pl_-X^J
-i\tilde{g}_{IJ}\tilde{\psi}^I_+D_+ \tilde{\psi}^J_+-i\tilde{g}_{IJ}\psi^I_-
\tilde{D}_- \tilde{\psi}^J_-+\right. \nonumber \\
&&\left.\frac{1}{2}\tilde{\psi}^I_+\tilde{\psi}^J_+\tilde{\psi}^K_-
\tilde{\psi}^L_-\tilde{R}_{IJKL}(\tilde{\Gamma}_+) \right)
\, . \label{sa}
\ee
The dual metric $\tilde{g}_{IJ}$ and antisymmetric field $\tilde{b}_{IJ}$ are
given by the Buscher's transformation \cite{bus}
and the dilaton shift has been omitted.
 In addition,  the fermions $\tilde{\psi}^I$ in the dual
theory  are related to the original ones  by
\bb
\tilde{\psi}^0&=&-\bar{\gamma}g_{0I}\psi^I+b_{0I}\psi^I \, , \label{psi0}\\
\tilde{\psi}^i&=&\psi^i \, , \label{pp}
\ee
where $\bar{\gamma}=\sigma^3$ is the corresponding $\gamma^5$ matrix in 2
dimensions.
This transformation is necessary in order the dual action, as the original
one, to has manifest N=1 supersymmetry \cite{1},\cite{bus}
and it can also be derived from the
supersymmetry transformations in eq.(\ref{sup}) \cite{2''}.

The transformation of the  fermion fields in eqs.(\ref{psi0},\ref{pp}) now,
effects the fermionic measure
$d\mu_f$ in the partition function. In particular, we found in \cite{1}
that the Jacobian of the transformation $\psi^I\rightarrow \tilde{\psi}^I$ is
the parity of the spin structure $(-1)^n$ where n is the number of (positive
chirality) zero
modes of  the Dirac operator. As a result, the original and the dual
theories are not the same but rather satisfy
\bb
Z[g,b,s]=(-1)^nZ[\tilde{g},\tilde{b},s] \, . \label{zzz}
\ee
Thus, for  odd world-sheet spin structures,
 the N=1 supersymmetric $\sigma$-model is not
invariant under T-duality. On the other hand, since  in  string  theory
 modular invariance requires a sum over all spin
structures, the string partition function cannot be invariant under T-duality
as well.
However, as we will see in a moment,
 T-duality maps type IIA to type IIB strings and vice
versa and in this sense it is a symmetry of type II theory. To
establish this, let us first consider  world sheet topology of a
2-torus which is the first non-trivial case where the anomaly shows up (since
$n=0$ for the 2-sphere).
In this case there
exist four spin structures labeled as ((+,+),(+,--),(--,+),(--,--))
which correspond to the possible
boundary conditions for the left- right-handed fermions on the torus
\cite{aaa},\cite{ws}.  The
partition function may then be written as a sum over all these structures as
\bb
Z[g,b]&=&\eta_{(+,+)}Z[g,b,(+,+)]+
\eta_{(+,-)}Z[g,b,(+,-)]+ \nonumber \\ &&
\eta_{(-,+)}Z[g,b,(-,+)]
+\eta_{(-,-)}Z[g,b,(-,-)] \, , \label{fz}
\ee
where the coefficients $\eta$  have to be specified by modular invariance
\cite{ws}.
It follows then from eqs.(\ref{zzz},\ref{fz}) that the partition function of
the dual theory is
\bb
Z[\tilde{g},\tilde{b}]&=&-\eta_{(+,+)}Z[g,b,(+,+)]+
\eta_{(+,-)}Z[g,b,(+,-)]+ \nonumber \\&&
\eta_{(-,+)}Z[g,b,(-,+)]
+\eta_{(-,-)}Z[g,b,(-,-)] \, , \label{ffz}
\ee
since only the (+,+) spin structure has a zero mode. This means that
$\tilde{\eta}_{(+,+)}=-\eta_{(+,+)}$. Thus for example,  if
$\eta_{(+,+)}=+1$ so that left- and right-handed fermions
are of the same chirality  corresponding to type   IIB string,
$\tilde{\eta}_{(+,+)}=-1$ in the dual theory and left- and right-handed
fermions are of the opposite chirality corresponding to type IIA theory and
vice versa \cite{ws}.

For higher genus Riemann surfaces now, the corresponding to eq.(\ref{fz}) sum
 will looks like
\bb
Z[g,b]=\sum_{s_+=1}^{2^{g-1}(2^g+1)}\eta_{s_+}Z[g,b,s_+]+
\sum_{s_-=1}^{2^{g-1}(2^g-1)}\eta_{s_-}Z[g,b,s_-] \, , \label{sff}
\ee
since there exist $2^{g-1}(2^g+1)$ even and  $2^{g-1}(2^g-1)$ odd  spin
structures on a genus g Riemann surface \cite{at}. Under T-duality, only the
second sum
will change sign since the parity for odd (even) spin structures is $-1(+1)$.
This means that the projections in the Ramond sector of left-handed and
right-handed fermions in the dual theory will be opposite to the original
theory turning a type IIA theory to type IIB and vice versa.

It should be noted that one has to consider global aspects of the
procedure as well \cite{3}.
 In particular, the holonomies $\int A$ of the gauge field
along non-trivial cycles of the world sheet  have  to be
integer numbers. This
specifies the range of the dual field $\tilde{X}^0$ which is restricted to lie
on the dual lattice of $X^0$. In this way the dual and the original theory are
also exactly equivalent as CFT. Finally, for an even number of abelian
isometries, T-duality is an exact symmetry, since each isometry will contribute
a $(-1)^n$ factor to the partition function leaving the latter invariant. Thus,
type IIA theory is T-dual to type IIB only for targets with odd number of
abelian isometries.

\vspace{.3cm}

I would like to thank E. Kiritsis and C. Bachas for a crucial   discussion.
This works supported in parts by the EC
programs  with contract no. SC1-CT91-0729 and SC1-CT92-0789.

\end{document}